\documentclass[letterpaper]{article} 
\usepackage{aaai25}  
\usepackage{times}  
\usepackage{helvet}  
\usepackage{courier}  
\usepackage[hyphens]{url}  
\usepackage{graphicx} 
\usepackage{natbib}  
\usepackage{caption} 
\usepackage{algorithm}
\usepackage{algorithmic}
\usepackage{multirow}
\usepackage{array}
\usepackage{dcolumn}

\usepackage{amsmath, amssymb}
\usepackage{booktabs}
\usepackage{newfloat}
\usepackage{listings}
\DeclareCaptionStyle{ruled}{labelfont=normalfont,labelsep=colon,strut=off} 
\floatstyle{ruled}
\newfloat{listing}{tb}{lst}{}
\floatname{listing}{Listing}
\usepackage{xcolor}

\title{The Role of Follow Networks and Twitter's Content Recommender on Partisan Skew and Rumor Exposure during the 2022 U.S. Midterm Election}

\author {
    Kayla Duskin,
     Joseph S. Schafer,
     Alexandros Efstratiou, 
     Jevin D. West,
     Emma S. Spiro
}
\affiliations {
    University of Washington, Seattle, Washington, USA\\
    \{kduskin, schaferj, aefstra, jevinw, espiro\}@uw.edu
}

\usepackage{bibentry}

\begin{document}

\maketitle

\begin{abstract}
Social media platforms shape users' experiences through the algorithmic systems they deploy. In this study, we examine to what extent Twitter's content recommender, in conjunction with a user's social network, impacts the topic, political skew, and reliability of information served on the platform during a high-stakes election. We utilize automated accounts to document Twitter's algorithmically curated and reverse chronological timelines throughout the U.S. 2022 midterm election. We find that the algorithmic timeline measurably influences exposure to election content, partisan skew, and the prevalence of low-quality information and election rumors. Critically, these impacts are mediated by the partisan makeup of one's personal social network, which often exerts greater influence than the algorithm alone. We find that the algorithmic feed decreases the proportion of election content shown to left-leaning accounts, and that it skews content toward right-leaning sources when compared to the reverse chronological feed. We additionally find evidence that the algorithmic system increases the prevalence of election-related rumors for right-leaning accounts, and has mixed effects on the prevalence of low-quality information sources. Our work provides insight into the outcomes of Twitter's complex recommender system at a crucial time period before controversial changes to the platform and in the midst of nationwide elections and highlights the need for ongoing study of algorithmic systems and their role in democratic processes.
\end{abstract}


\section{Introduction}

Social media, and the recommendation algorithms that power them, play a critical role in information dissemination during high stakes events such as elections \cite{stier2020election, Tucker2018-lj}. The speed of communication, the ability for candidates to communicate directly with the public, and the opportunity for users to engage with one another make social media appealing for both communicators and consumers of political and election information. 
The social media platform Twitter/X\footnote{The platform has been renamed to X. Given that this study was conducted prior to that change we will refer to the platform as Twitter for the remainder of this manuscript.} is particularly known for supplying users with political news, with 74\% of users in the United States (U.S.) saying they see political content on the platform \cite{pew_politics_2024}. 

However, social media as a medium for civic information is not without drawbacks. The challenges and consequences of online rumors and mis/disinformation have become especially salient in recent election cycles, particularly in the U.S. \cite{Sharma2022-zx, Kennedy2022-qv, Green2022-ct}. Additionally, users are often surrounded in their online community by those with similar worldviews and perspectives \cite{Nyhan2023-st, Cinelli2021-jr}, though not more so than offline \cite{Ross-Arguedas2022-ep}. There is concern that encountering news and information through social media may offer a distorted perspective of the political landscape and highlight the most extreme views in a way that is detrimental to social cohesion \cite{Gonzalez-Bailon2023-nz}.

Users' experience of social media, including their exposure to critical civic information, is mediated by the recommendation algorithms implemented by the platform. We know that these algorithms increase the amount of time a user spends on the platform and affect how users engage with posts~\cite{Guess2023-ia, Wang2024-zd}. However, much less is understood about these algorithms' influence during a high stakes election or how this compares to the influence of a user's chosen network of friends.

To shed light on this subject, we deploy automated Twitter accounts with varying social network compositions and systematically document the content served to them during the time period surrounding the 2022 U.S. midterm elections. Crucially, the use of automated accounts allows for analysis of the algorithmically curated timelines personalized for each account alongside the corresponding reverse chronological timeline comprised of posts by those within their social network. Taking into account the social context of the accounts, we describe the role of Twitter's content recommender surrounding a major election, when accurate and timely information is crucial.

Specifically, in this study we seek to answer:

\begin{itemize}
    \item \textbf{RQ1}: What is the effect of the recommender on the volume and type of election content served to users?
    \item \textbf{RQ2:} Does the recommender influence the partisan skew of information, and if so, how?
    \item \textbf{RQ3:} Does the recommender impact information quality, through exposure to unreliable sources and false rumors, and if so, how?  
\end{itemize}

We find that the partisan makeup of one's network heavily influences the content experienced in both feeds, and that the recommendation algorithm had a measurable effect on the volume and source of election content, the partisan skew of featured authors, and the frequency of exposure to election-related rumors and low-quality links. Critically, we find that algorithmic influence varies depending on the partisan makeup of one's interpersonal network.

The algorithmic feed showed less election content to left-leaning accounts only, but increased how often accounts of any partisanship saw tweets by political candidates running for office. The partisan makeup of each account's friend network played the largest role in predicting the political slant of timeline tweets, but the algorithmic feed contributes a small right-leaning shift for all groups. Recommended content was more likely to contain tweets related to false or misleading election narratives for right-leaning and neutral accounts. Finally, while accounts saw fewer tweets that link to low-quality information sources \textit{overall} on their algorithmic feed, the content recommender increased the relative frequency of low-quality sources for left-leaning and neutral accounts.  

\paragraph{Scope and contributions.}
By analyzing content served to live accounts, our analysis captures the final product of algorithmic design when combined with real-world events and network structure. While simulations based on recommendation system code can be useful in understanding the \textit{possible} impacts of algorithmic design, knowledge of the source code is insufficient to understanding its \textit{realized} impacts given highly complex contexts \cite{Rahwan2019-yl, atlanticTransperent, wiredTwittersOpen}. \citet{Ribeiro2023-ti} point out that algorithmic audits using automated accounts may inadequately capture amplification effects if not accounting for user preferences (e.g. internal preference for what video to watch next on YouTube, separate from what is recommended). In this study, we focus on algorithmic behavior given baseline social network preferences instead of amplification given user behavior. This offers an estimate of the baseline, or \textit{algorithmic priors} of the recommender in the absence of content engagement feedback.

These findings are critical in documenting the ever-changing landscape of online political discourse and the systems that mediate that discourse. Importantly, we provide an in-situ study on Twitter during a major U.S. election and in a time period preceding major changes in Twitter's operation. This study establishes a baseline from which future deployed recommender systems may be measured --- a critical need given the recent volatility in platform governance \cite{Myers2023-al, pew_politics_2024}.

\section{Related Work}
The role content recommenders play in online information ecosystems has been under increasing scrutiny as it becomes clear that algorithmic systems have the power to influence user experiences, information availability, and social interactions. In this section, we briefly summarize the concept of algorithmic audits, prior studies of Twitter's content recommender in particular, and the interplay between algorithmic recommendation and U.S. elections. 

\subsection{Algorithmic Audits}
Internet users' experiences are increasingly mediated by sophisticated algorithms that make automated choices on what kinds of content, social connections, and information is easily accessible to users \cite{Rahwan2019-yl}. Many of these systems are proprietary and opaque, leaving the public to guess if and how these invisible decisions bias their experience. The concept of algorithmic accountability, which goes beyond transparency by assessing the impact of black box systems, has been proposed as a crucial topic of research given the degree of power these systems may exert \cite{Diakopoulos2015-it}. By observing the outcomes of an algorithmic system under various conditions, algorithmic audits can reveal patterns of bias or influence \cite{Sandvig2014-er,Metaxa2021-ex}. While there are several ways to conduct algorithmic audits, in this study, we employ automated accounts, or ``sock puppets'', to emulate the experience of a user while controlling for relevant variables, a method useful in understanding algorithmic influence \cite{Diakopoulos2021-dq}.

\subsection{Content Recommendation on Twitter}
Twitter's content recommendation algorithm (used to populate users' `Home' timeline) has been studied by those seeking to understand its influence on emotion-provoking content, news media, and information diversity. Given that the recommendation algorithm has likely changed over time, it is difficult to directly compare outcomes. The works described here are notable for their methodological contributions and provide important snapshots of the platform. Only through repeated studies of this dynamic system are we able to identify trends or emerging issues.

Concern over the amplification of emotionally-charged or insensitive content has been a core driver of content recommendation research. In a randomized experiment of users assigned to the reverse chronological feed rather than the algorithmic feed, \citet{Bandy2023-xb} determined that Twitter's algorithmic ranking exposed users to lower rates of marginally abusive (otherwise known as toxic) content, but that users experiencing the algorithmic ranking consumed more toxic content in sum. In their experimental work conducted with participants using a custom browser extension, \citet{Milli2025-lv} found that the platform's recommender did not silo users into ideological `filter bubbles', but that it did amplify emotionally-charged content that was more likely to denigrate a user's political out-group.

In terms of information diversity, Bandy and Diakopoulos (2021) leveraged automated accounts to assess the impact of Twitter's algorithmic curation and found that while the content recommender increased the diversity of content sources, it suppressed bipartisan accounts and amplified more niche political accounts along partisan lines \cite{Bandy2021-lg}. Other research using automated accounts has shown that Twitter's algorithmically curated timeline favors more popular tweets, and skews which friends users see content from \cite{Bartley2021-we}. 

Another thread of research explores the impact of Twitter's content recommender on news media and politics. In a study of participants utilizing browser extensions,  \citet{Wang2024-zd} conducted a three week experiment, finding that Twitter's algorithmic timeline exposed participants to less content from news sources, but that the sources are less extreme and marginally more trustworthy. In contrast, \citet{Bandy2021-dl} found that both major news brands and junk news websites received more exposure share on the algorithmic timeline, though overall frequency of news was decreased, perhaps due to conducting the study in a different time period. In one study, researchers within Twitter found through a randomized experiment that, on aggregate, the personalized home timeline algorithm amplified right-leaning politicians and media outlets \cite{Huszar2022-jm}. In this study, we contribute to the understanding of content recommendation in the specific context of the 2022 U.S. midterm elections, conducting an audit external to the platform which corroborates the findings of \citet{Huszar2022-jm}.

\subsection{Algorithmic Influence in Election Contexts}
Facebook's collaboration with academic researchers to study the 2020 U.S. presidential election is one of a limited number of studies investigating how algorithmic systems influence user experiences during major elections, something the research community has called for increased attention to investigate \cite{mustafaraj_case_2020}. Through a randomized experiment of Facebook users, \citet{Guess2023-ia} found that the algorithmically curated feed on the platform increased the proportion of content from ideologically like-minded sources while decreasing the proportion of content from untrustworthy sources when compared to a reverse chronological feed. However, this decrease in exposure to untrustworthy sources may have been the result of a temporary measure implemented by the platform \cite{Bagchi2024-hz}. Another study conducted during the 2020 U.S. election utilized internet browsing data to estimate the impact of changes to Facebook's News Feed algorithm \cite{Bandy2023-hc}. They found that as a whole, Facebook amplified the reach of low-quality publishers, but that changes made to the recommendation system did have an effect, lowering visits to low-quality sources. \citet{Duskin2024-sy} conducted an audit of Twitter's `Who to Follow' recommendation algorithm during the 2022 U.S. midterm elections, finding that the friend recommendation algorithm constructed networks that were less politically homogeneous and less likely to share false or misleading election narratives than networks that were built based on social endorsement. Most similar to our work is that of \citet{Ye2025-ej}, conducting a sock puppet audit on Twitter during the 2024 U.S. presidential election, focusing specifically on how the recommender (de)amplifies out of network users. Our studies differ in that \citet{Ye2025-ej} focus on \textit{user-level} amplification and restrict their automated accounts to following media accounts and political entities. Our study focuses on content-level analysis with an emphasis on information quality, and populate the networks of the automated accounts without restriction on the type of account. Our findings corroborate theirs in the large impact of one's local social network, and we find a small right-leaning skew in recommended content which they also identify for accounts that do not follow anybody (``neutral" accounts).

\section{Data \& Methods}
\subsection{Audit Accounts}
Twitter's API has never made data from users' algorithmically curated timeline available, creating a challenge for those interested in studying its impacts. We make use of automated accounts, which we refer to as audit accounts, to enable the comparison between content displayed on Twitter's reverse chronological and algorithmically curated timelines; this approach has been used in prior work facing similar data access challenges. The use of automated audit accounts has several advantages over other research methods (e.g. simulations). The most salient is the ability to observe the cumulative, real-world, effect of the many complex factors that influence online information environments in response to account characteristics and interactions \cite{Diakopoulos2021-dq}. While the recent publication of the code for Twitter's recommender system\footnote{\url{https://github.com/twitter/the-algorithm}} is a step toward transparency, it does not provide comprehensive insight into its impacts, given that recommendations are contingent on the complex and largely opaque social network underlying the platform \cite{atlanticTransperent,wiredTwittersOpen,Rahwan2019-yl}. An in-situ audit is one way to explore the cumulative outcomes of Twitter's system as it operates amid ever-changing current events and shifting social connections. 

\subsubsection{Account Creation}
Audit studies of networked platforms employing automated accounts rely on a variety of strategies for initiating the social ties of the audit accounts. Some have constructed networks by following a random sample of pro-science and anti-science Twitter accounts \cite{Bartley2021-we}, while others have attempted to create archetypal left and right partisans \cite{Bandy2021-dl,Ye2025-ej}. Another approach uses stochastic following behavior to form networks after seeding an `initial friend'  \cite{Chen2021-tg}. In this study, we employ 20 automated accounts using a method of stochastic network growth after initiating each account with a single seed ``friend'' chosen from candidates running for U.S. Senate. We describe this approach in the following sections, and further technical detail for the method we follow can be found in \citet{Duskin2024-sy}.

To start, the research team identified U.S. Senate races that were predicted to be close races in 2022\footnote{These were identified through the popular poll aggregation site \url{https://projects.fivethirtyeight.com/2022-election-forecast/senate/}} and set five Democratic candidates and five Republican candidates (opponents in five Senate races) as the seed accounts. We then created 20 Twitter accounts and initiated each with a single ``friend'' such that two unique accounts followed the account of each of the ten identified Senate candidates. We subsequently grew these networks stochastically, but intentionally set the seeds to be balanced in number between Democrats and Republicans and directly connecting to political figures (rather than news media).
Following their initiation, each of the audit accounts expanded their network in one of two ways, resulting in one audit account following each method for each of the ten seed friends. To accumulate social ties, half of the audit accounts followed users that were retweeted by their current friends: twice a day we retrieved the most recent 200 retweets by accounts already in their network and then randomly sampled six accounts that the audit account would add as a friend by following. The other half of the audit accounts followed users recommended by Twitter's `Who to Follow' suggestions: twice a day we retrieved six accounts recommended to the audit account using Selenium to scrape the `Who to Follow' which is displayed in the lower right of the Twitter home page for logged in users. In this design, all audit accounts steadily grew their networks in ways that reflect how new users to the platform may expand the accounts they follow. This network growth process was conducted starting in September, 2022 and continued through November, 2022. Timeline data for two of the original 20 accounts was incomplete due to restrictions placed on the audit accounts by Twitter; we dropped those observations and analyze data from the remaining 18 accounts.

\begin{figure}[t]
\centering
\includegraphics[width=\columnwidth]{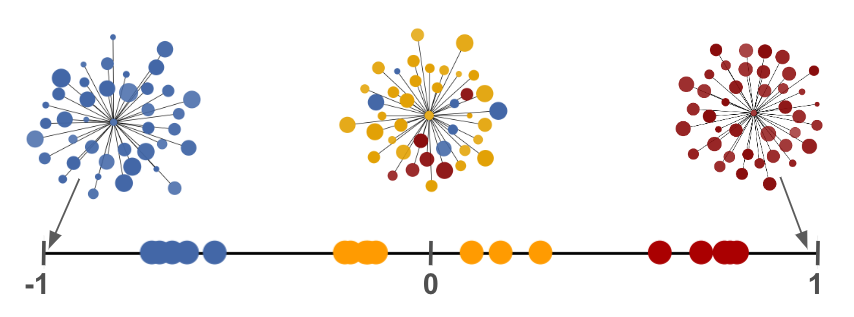}
\caption{\small Audit accounts, plotted according to average network partisanship, with dots colored according to their network-leaning classification (blue for left-leaning, yellow for neutral, red for right-leaning). A score of -1 would indicate that every account they follow is left-leaning, while 1 would indicate that every account they follow is right-leaning. Network visualizations are idealized examples of what networks at key partisanship levels might look like.}
\label{fig:network_leaning}
\end{figure}

\subsubsection{Account Behavior}
The automated accounts are passive consumers of information --- they do not post, repost, or like content, or engage with platform users outside of following other accounts. Though Twitter uses on-platform actions as input to its content recommender, we avoid introducing content engagement actions for three reasons. First, this choice acts to isolate the specific role of network composition within Twitter's recommendation algorithm --- separate from the concept of an algorithmic feedback loop, or ``rabbit hole", where algorithmic recommendations and user feedback interact. This phenomenon has been studied on YouTube \cite{ribeiro_auditing_2020, Hosseinmardi2024-vn}, and merits future study on other platforms but here we focus on identifying the ``priors" of the content recommender based solely on the social network. Second, simulating engagement actions would introduce further variance between accounts that may confound results.
Finally, given the fraught nature of U.S. elections in recent years, we could not ethically justify the risk of unintentionally amplifying, endorsing, or sharing misleading or divisive content surrounding the election. Furthermore, this behavior is not out of the norm for typical users of the platform, as the bottom 75\% of Twitter users produce a median of 0 posts (including original tweets, retweets, and replies) per month \cite{McClain2021-ku}.

\subsubsection{Account Network-leaning}
Twitter's content recommender takes into account a user's network of friends (as well as second degree connections, or \textit{friends of friends}) in determining content to serve.\footnote{\url{https://blog.x.com/engineering/en_us/topics/open-source/2023/twitter-recommendation-algorithm}} Since the audit accounts do not engage with other users (beyond following them) or interact with content on their feeds, this study isolates the interaction effects between the network of accounts followed by the audit account and the content recommendation algorithm. While starting from either a single Republican or Democratic Senate candidate, the resulting social networks of the audit accounts (i.e. their friends) after stochastic growth emerge as ideologically diverse.

To capture the ideological makeup of each audit account's local social network, we construct a categorical variable with three levels to capture the partisan makeup of each account's immediate neighborhood on the platform, along the left (liberal) to right (conservative) axis of American politics.\footnote{The labels left-leaning and right-leaning are used here within the context of the U.S. political landscape and may differ from how these terms are used internationally.} We define categorical \textit{network-leaning} for each account based on the average political behavior of the accounts they follow. To do this we first calculate the partisanship of each of the accounts they follow (see \textbf{Partisan Content} section), and take the mean so that each of the audit accounts has an average network partisanship score on a scale between -1 (entirely left-leaning friends) and 1 (entirely right-leaning friends). 

Some outcomes of interest do not make sense along the raw, monotonic scale of average network partisanship (e.g. accounts with average network partisanship near zero likely see less elections content than those with larger absolute network-leanings), making categorical labels more suitable. The highly clustered distribution of scores, as shown in Figure \ref{fig:network_leaning}, provides intuitive cut-offs for translating average network partisanship to three levels. We assign each account's network-leaning, $l$, based on the average partisanship, $\bar{p}$, of their friends: 
\[
l=\begin{cases} 
      \text{Left-leaning} & \phantom{-}\bar{p} \leq -0.5 \\
      \text{Neutral} & -0.5 \leq \bar{p} \leq 0.5 \\
      \text{Right-leaning} &  \phantom{-}0.5 \leq \bar{p} 
   \end{cases}
\]

As a result, five accounts are categorized as left-leaning (mean network partisanship: -0.65) , five accounts are categorized as right-leaning (mean network partisanship: 0.72), and eight accounts are categorized as neutral (mean network partisanship: -0.04).

\subsection{Timeline Tweets}
In order to assess the influence of Twitter's content recommendation algorithm, we collected the tweets served to each audit account on their `Home Timeline', also called their \textit{algorithmic feed}, and their `Following' timeline, which we refer to as the \textit{chronological feed} twice during twice-daily `sessions' between November 2nd, 2022 and November 25th, 2022. During this observation period, every account logged in twice daily for a session once between approximately 6am and 9am PST and once between approximately 6pm and 9pm PST. Times within the morning and evening windows were randomized across accounts. During each session, the IDs and content of the first 50 tweets from the account's algorithmic feed and the first 50 tweets from the account's chronological feed were collected. We did not include non-tweet content such as ads in the collection of either timeline. We collected this data by implementing a scheduled Python script and used Selenium\footnote{\url{https://selenium-python.readthedocs.io/}} to conduct the automated log-ins and content scraping. Once per day, we used the IDs from the observed timeline tweets to collect (i.e. `rehydrate') the full tweet information from Twitter's V2 API.

\subsection{Data Enrichment}
To understand the kinds of content shown to the audit accounts, we annotated the dataset of timeline tweets. For each tweet served in the chronological or algorithmic timeline, we identified whether the tweet pertains to the midterm election, whether it was authored by a politician, whether it referenced a false or misleading election narrative, the estimated political leaning of its author, and the quality of the domain in any included URL. Here we detail the methods used to operationalize these features. 

\subsubsection{Election Content}
\label{sec:election_content}
We annotate each tweet's relevance to the political conversations surrounding the election. To do so, we cross-reference timeline tweets with a dataset of 446 million tweets collected during the 2022 midterm election described in \cite{Schafer2025-fh}. This dataset was built using Twitter's streaming API, querying based on a set of election-related keywords developed by subject matter experts (terms like ``election", and ``ballot" alongside more generally political terms such as ``republican" and ``democrat".)
We describe our procedure for validating this annotation and provide examples in the Appendix.

\subsubsection{Elected Officials and Political Candidates}
Individuals holding or running for public office often use Twitter to comment on key issues and raise support with the public \cite{Shah2022-wm, Devlin2020-ex}. We annotate tweets authored by a politician, which we define as those holding office or running for office during the study time period. To identify content shared by political candidates, we used a list of the Twitter handles of candidates running for U.S. political office in 2022. This list was obtained by first scraping Wikipedia's list of candidates running in senate, congressional, or gubernatorial races.\footnote{\url{https://wikipedia.org/wiki/2022_United_States_Senate_elections}, \url{https://wikipedia.org/wiki/2022_United_States_House_of_Representatives_elections}, \url{https://wikipedia.org/wiki/2022_United_States_gubernatorial_elections}}

Then, using this list of candidates, Twitter handles (when existent) were scraped from \texttt{ballotpedia.org}. We identified content posted by sitting officials by subsetting data published by \texttt{the @unitedstates project} which shares information on U.S. legislators including social media handles.\footnote{https://github.com/
unitedstates/congress-legislators.} We combine the lists of candidate and elected official Twitter IDS to form our reference set of politicians. For the present study, we cross-referenced this list of politicians' Twitter user IDs with the authors of the posts in our timeline tweets dataset. This process identifies all tweets found on the timelines of the audit accounts that were authored by politicians; note, these posts may be directly related to the election or may offer the candidate's take on current events and issues.

\subsubsection{Partisan Content}
\label{sec:partisanship}
We measured the impact of the recommender on partisan skew via the magnitude and direction of partisanship of content \textit{sources}. We consider content sources in two ways: via the estimated partisanship of tweet authors and via the Media Bias Fact Check (MBFC) ``bias score'' of external domains included as links.

We have described the process for assigning each of the audit accounts a \textit{network-leaning} category in the \textbf{Account Network-leaning} section above. Here, we describe the process for estimating the partisanship of other accounts on the platform. To estimate the partisanship of tweet authors (i.e. accounts), we use the method based on partisan participation detailed in \citet{Schafer2025-fh}. In brief, this is done by clustering accounts engaged in election discourse using a coengagement network to identify left and right leaning influential accounts \cite{Beers2023-qe}. 
For accounts not present in the influential user clusters, partisanship scores are assigned according to which influential accounts they have retweeted. A score of -1 indicates that they primarily (more than 80\% of the time) retweet left-leaning influencers, a score of 1 indicates that they primarily retweet right-leaning influencers, and a score of 0 indicates that they either do not engage with political influencers or equally retweet from both sides. Some popular methods for estimating partisanship rely on following relationships to infer partisanship \cite{Barbera2015-og,Mosleh2022-aj}. We rely on retweeting behavior as a signal of ideological homophily for two reasons. The first is that the collection of the full following networks for all accounts present in the timeline tweets data is computationally intractable given Twitter API rate limits. The second reason is that our method represents the partisanship \textit{enacted} by users active at the time of the study. That is, our method defines partisanship based on a user's active support (through retweets) of political influencers during the time period of interest.
For each observed session we take the mean partisanship score of tweet authors to indicate the overall slant of information sources.

To assess the skew of external sources, we rely on ``bias scores'' reported by Media Bias Fact Check (MBFC), an independent website that evaluates the political skew and factual reliability of news sources based on content analysis and adherence to journalistic standards.\footnote{\url{https://mediabiasfactcheck.com/}} Adapting data and code from \citet{sanchez2024mapping}, we identify the MBFC pages that exist for domains present in the timeline tweets data and scrape the pages to retrieve the bias score reported by MBFC. These scores are on a scale from -10 (extreme left bias) to 10 (extreme right bias). MBFC provides a bias \textit{category} for every domain they assess,\footnote{The categories are extreme left, left, left-center, least-biased, right-center, right, and extreme right} but they only provide \textit{scores} for some sites. We were able to identify 449 external domains from our data that had been rated by MBFC. Of the 449 domains, 180 (40\%) were provided numerical bias scores by the site. Bias scores for bias each category fall within mostly non-overlapping ranges (see Appendix). To obtain numerical scores for all 449 rated sites, we impute the bias scores for the remaining domains. Each domain is rated with a bias category as well as a ``factual reporting" category, and we impute the missing scores to be the bias score average for the domain's corresponding bias and factual category. 

\begin{figure*}[t]
\includegraphics[width=1.5\columnwidth]{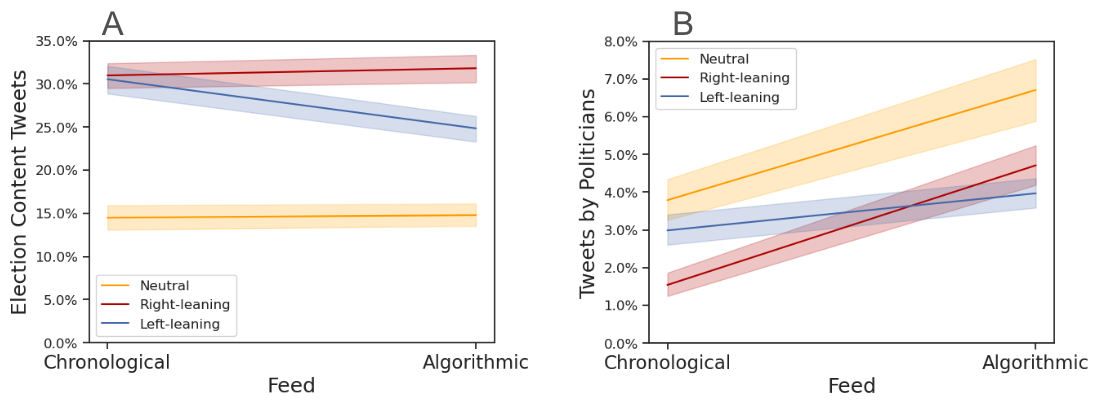}
\centering
\caption{\small Interaction effects between network-leaning and feed on the prevalence of (A) election content and (B) the prevalence of content posted by political candidates or elected officials.}
\label{fig:political}
\end{figure*}

\begin{table}[t]
\caption{Examples of tweets in our dataset that either directly reference a rumor about the election, or link to a domain labeled as low-quality.}

\renewcommand{\arraystretch}{1.2}
\begin{tabular}{m{3.85cm}m{3.85cm}}
\hline
\textbf{Election Rumor} & \textbf{Low-Quality Source}    \\ \hline
\footnotesize Nevada voted for a Republican governor yet re-elected a Democrat senator? I'm not buying it.  & \footnotesize Did Ukraine Try To Lie Us Into WWIII? - today on the Liberty Report: \textless{}link to site\textgreater{} \\ \hline
\footnotesize The Views Sunny Hostin openly admits to voter fraud. Where is the FBI on this one? & \footnotesize France Takes Just 38 Redistributed Migrants After Promising to Take 3,500 \textless{}link to site\textgreater{}                            \\ \hline
\footnotesize ``We will not allow Katie Hobbs to certify her own stolen election" - Gregg Phillips ... Arizona, America and the World is watching.               & \footnotesize In This Untold Story Of Poll Worker Data, Chinese Servers, And Scandal, Only The FBI Knows The Truth  \textless{}link to site\textgreater{}                    

\end{tabular}
\label{table:tweet_examples}
\end{table}

\subsubsection{Information Quality}
To assess the quality of information, we consider the presence of misleading election rumors as well as the quality of linked domains, with examples of both labeling methods in Table \ref{table:tweet_examples}. To annotate each tweet as related to a false, misleading, or uncorroborated narrative, we use the \textit{ElectionRumors2022} dataset \cite{Schafer2025-fh}. This dataset is a subset of the data on election-related discourse and contains approximately 1.8 million tweets related to specific false, misleading, or unsubstantiated election administration and election process rumors. Tweets are not individually annotated as false or misleading, but rather as \textit{pertaining to a false, misleading, or uncorroborated narrative} based on their inclusion in the \textit{ElectionRumors2022} dataset. 

We also annotate the quality of information through the application of external domain-level labels, as has become a common practice in prior work on information quality \cite{Mosleh2022-aj, Moore2023-wy,Baribi-Bartov2024-hn,Guess2023-ia}. We identify URLs present in each timeline tweet, resolving shortened URLs and retrieving the domain of each using the \texttt{requests} library.\footnote{\url{https://pypi.org/project/requests/}} Next, we cross-reference the identified domains to the Iffy Index of Unreliable Sources,\footnote{\url{https://iffy.news/}} which lists unreliable or low-quality information sources based on credibility ratings by Media Bias Factcheck. We annotate each tweet as containing a link to a low-quality source or not.   

\subsection{Analytic Approach}
We are interested in quantifying the effect of the recommended feed on content exposure in comparison to the reverse chronological baseline. To do so, we fit a separate multiple regression model for each outcome variable of interest. Specifically, we use mixed effects models, which allows us to analyze the within-subjects design of this experiment. That is, we can compare the outcomes of the algorithmic feed to the outcomes of the chronological feed \textit{within each audit account}, so each account acts as its own control. In each model we include feed type as a fixed effect with two levels --- algorithmic or chronological; this is our primary variable of interest. We also include a fixed effect with three levels to account for the network-leaning of each account and an interaction term between the feed type and network-leaning. We also include random effects that account for individual account variance across repeated measures. In each model we include account ID, session number, and a binary indicator of whether the account grew its network through algorithmic recommendation or based on friends' retweets.

When assessing author partisanship and domain skew scores, we use a linear mixed effects model. For all other outcomes of interest, we use a generalized linear mixed effects model with a logit link function to capture the binomial distribution of the binary response variables. All models are fit on 1728 observations from (18 accounts) x (48 sessions) x (2 feeds) which represent the 90,544 tweets seen across all observations. When the primary mixed effects model indicated significant interactions between fixed effects, post-hoc pairwise comparisons were conducted using Z-tests, with p-values corrected for multiple comparisons with Holm's sequential Bonferroni procedure.

\section{Results}

\subsection{RQ1: Election Content}
To assess the impact of the recommender on election content served to users surrounding a major election, we consider the proportion of content during each session that was determined to be election-related and how frequently tweets posted by a politician are shown. We find that election-related tweets made up an average of 27.7\% (95\% CI: 26.59-28.81\%) of the content seen by left-leaning accounts, and 31.4\% (95\% CI: 30.32-32.48\%) of content seen by right-leaning accounts, which is significantly more than accounts with neutral networks which saw 14.64\% (95\% CI: 13.68-15.6\%) election content ($\beta=-1.33, SE=0.43, p<0.01$). Interaction effects between the feed and the network-leaning can be seen in Figure \ref{fig:political} A, and these interactions were confirmed to be statistically significant via the mixed effects model. Post-hoc tests indicate that the algorithmic feed decreased the proportion of election-related content shown to left-leaning accounts ($Z=-10.64, p<.001$) from an average of 30.54\% (95\% CI: 28.95-32.14\%) of posts to 24.85\% (95\% CI: 23.4-26.31\%) while we did not find evidence for an effect for neutral or right-leaning accounts. 

We also find that the algorithmic feed increased the prevalence of content authored by politicians across the board. On average, 2.94\% (95\% CI: 2.65-3.23\%) of posts on the chronological feeds were authored by candidates, while 5.39\% (95\% CI: 4.98-5.81\%) of posts on the algorithmic feeds were authored by candidates. The difference between algorithmic and chronological feed is statistically significant for all network-leaning groups (left-leaning: $Z=4.18, p<.001$, neutral: $Z=13.37, p<.001$), but the effect was largest for right-leaning accounts ($Z=13.70, p<.001$). Note that this does not contradict, but does add nuance to findings about the prevalence of election content. Content related to elections may come from a variety of sources, including news outlets or the general public, and content recommendation may uniquely influence different sources leading to a reduction in overall election-related content while increasing visibility of content posted by candidates.

\subsection{RQ2: Partisan Skew}
\begin{figure}[!htb]
     \centering
     \includegraphics[width=.75\linewidth]{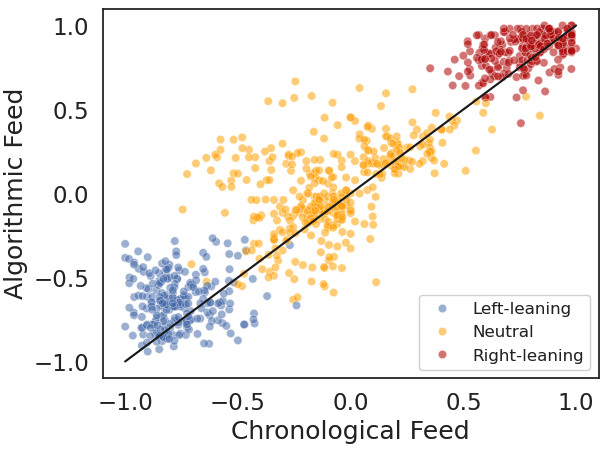}
     \caption{\small Average partisanship score of tweet authors displayed on the chronological feed and algorithmic feed. Points represent one session, and are colored based on the network-leaning of the viewing account. Points above the y=x line indicate sessions where the algorithmic feed was more right-leaning than the chronological feed and points below the line indicate sessions where it was more left-leaning.}\label{fig:author_scatter}
\end{figure}

The prospect that a content recommendation algorithm could unequally amplify different ideological perspectives is concerning, both regarding the possibility for amplifying extremes on both sides of the political spectrum, and the possibility of consistently amplifying one side over the other. In our data, we see evidence for the latter, though this effect is comparatively small compared to the impact of the partisan makeup of one's network, visually shown in Figure \ref{fig:author_scatter}. We find statistical backing for this, as the estimated coefficients within the mixed effects model are positive (indicating a more right-leaning average partisanship) and statistically significant for the effect of the algorithmic feed when compared to the chronological feed ($\beta=0.12, SE=0.01, p<.001$), as well as for the effect of neutral network-leaning ($\beta=0.69, SE=0.09, p<.001$) and right network-leaning when compared to left-leaning ($\beta=1.54, SE=0.1, p<.001$).

\begin{figure}
    \centering
    \includegraphics[width=.8\linewidth]{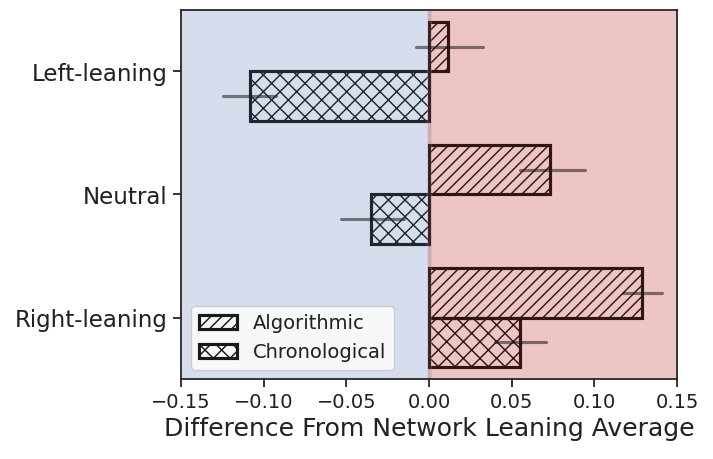}
    \caption{\small Mean difference between the average partisanship of authors displayed on the chronological and algorithmic feeds, and the average partisanship of an account's friend network. These differences, except for the difference in algorithmic feed for left-leaning accounts, were all found to be significant based on paired samples t-testing using Holm’s sequential Bonferroni procedure to correct for multiple comparison.}
    \label{fig:author_bar}
\end{figure}

Interestingly, we see that the average author partisanship of both the algorithmic feed and the chronological feed vary from the average partisanship of accounts' friend networks. Figure \ref{fig:author_bar} shows that for neutral and left-leaning accounts, their chronological feeds were significantly more left-leaning than their social network. The algorithmic feed for neutral accounts was significantly more right-leaning, while both the chronological and algorithmic feed for right-leaning accounts was more right-leaning than their network.  

We also consider the skew of external sources of information, based on the bias ratings of the external sites that accounts are exposed to via links. We find that the algorithmic feed shifts the skew experienced by the neutral accounts slightly to the right, from an average of -0.44 (95\% CI: -0.84- -0.04) to an average of 2.16 (95\% CI: 1.58-2.75). ($t=8.46, p<.001$). The recommended feed also shifts the skew experienced by right-leaning accounts very slightly to the right, from an average of 4.63 (95\% CI: 4.32-4.94) to 5.49 (95\% CI: 4.8-6.16) ($t=2.81, p<.05$). These shifts are not large enough to equate to a meaningful rightward shift --- the neutral accounts experience `least-biased' sources on average while right-leaning accounts experience `right-center' sources on average, regardless of feed.

\subsection{RQ3: Information Quality}
\begin{figure}[t]
\centering
\includegraphics[width=\columnwidth]{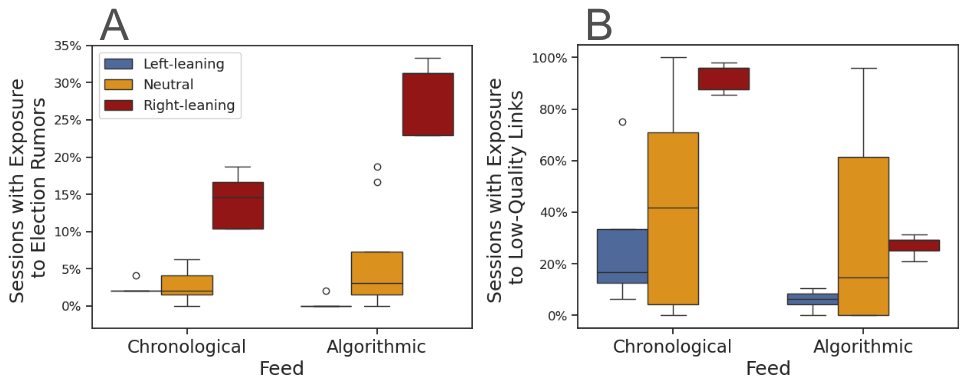}
\caption{\small Box plots showing the proportion of sessions where each account is exposed to a false or misleading narrative (A), or a link to a low-quality external domain (B). Note that A and B do not share a y-axis.}
\label{fig:misinfo_box}
\end{figure}

Our findings on exposure to election rumors and low-quality information corroborate established findings on partisan asymmetry in exposure to misinformation on social media \cite{Mosleh2022-aj, Gonzalez-Bailon2023-nf, Madani2025-po}. We additionally probe how the algorithmic recommendation interacts with partisan differences. To assess the influence of Twitter's content recommender on the quality of information that users are exposed to in feeds, we consider two methods of identifying low-quality information. The first considers individual tweets which reference false, misleading, or uncorroborated election narratives. The tweets identified here come from a high-precision dataset of 135 unique election-related rumors circulating during the 2022 U.S. midterm election, such as the rumor the using a sharpie pen will invalidate your ballot \cite{Schafer2025-fh}. Figure \ref{fig:misinfo_box} A shows the proportion of sessions in which accounts encountered at least one of these rumors and the large difference between accounts with right-leaning networks and the other network-leaning groups ($\beta=2.2, SE=0.6, p<.001$). On average, left-leaning accounts encountered one of these narratives on their chronological feed during 2.5\% (95\% CI: 0.52-4.48\%) of sessions and on their algorithmic feed during 0.42\% (95\% CI: -0.4-1.23\%) of sessions. Neutral accounts saw them moderately more frequently with 2.6\% (95\% CI: 1.01-4.2\%) of chronological sessions and 5.99\% (95\% CI: 3.61-8.37\%) of algorithmic feed sessions containing election rumor narratives. In contrast, accounts with right-leaning networks saw at least one false or misleading election narrative during 14.17\% (95\% CI: 9.75-18.59\%) of sessions on their chronological feed and 28.33\% (95\% CI: 22.62-34.05\%) of sessions on their algorithmic feed. Significant interaction effects between network-leaning and feed was established by the generalized mixed effects model. Post hoc tests did not identify significant impact for left-leaning or neutral accounts, but we did find evidence that the algorithmic feed \textit{increased} the frequency of exposure for right-leaning accounts (Z=4.21, p$<$.001). 

The second method of evaluating information quality is by identifying how frequently accounts were exposed to links to unreliable sources. In this case, we use a list of domains that frequently fail fact-checks. It is more common for accounts of any network-leaning to be exposed to at least one low quality link than tweets directly related to election rumors, as seen in Figure \ref{fig:misinfo_box}. Figure \ref{fig:misinfo_box} B also shows that for both left-leaning and right-leaning accounts, there was less exposure to low quality domains on the algorithmic feed, with left-leaning accounts encountering these domains during only 5.83\% (95\% CI: 2.86-8.8\%) of sessions on their algorithmic feed compared to during 28.75\% (95\% CI: 23.01-34.49\%) of sessions on their chronological feed. Right-leaning accounts encountered a low quality domain during 27.08\% (95\% CI: 21.45-32.72\%) of their sessions on the algorithmic feed, compared to 92.5\% (95\% CI: 89.16-95.84\%) of chronological feed sessions. There is a smaller relative difference for neutral accounts: the algorithmic feed reduced exposure from 43.75\% (95\% CI: 38.78-48.72\%) of chronological sessions to 32.29\% (95\% CI: 27.61-36.97\%) of algorithmic sessions.   

While the method of classifying information quality based on its source rather than its contents is frequently used, there are drawbacks to this approach; namely that only tweets containing links to external sources can be identified as low-quality. Prior work has shown that Twitter's algorithmic feed displays significantly fewer tweets with links to external sources when compared to the chronological feed \cite{Bandy2021-lg, Wang2024-zd, Bandy2021-dl}. Our study corroborates these findings: the algorithmic feed strongly reduced the proportion of external links ($\beta=-3.02, SE=0.08, p<.001$) and there was no detectable difference for exposure to external links across network-leaning groups. For this reason, the raw proportions of exposure to low-quality sources presented in Figure \ref{fig:misinfo_box} B show the compounded effect of the algorithm's reduction of total links with any possible effect of the recommender specific to low-quality links. 

To disentangle the effects of the recommendation algorithm on the prevalence of external links and the prevalence of links to low-quality sites, we directly analyze the impact of the algorithmic feed on the prevalence of low-quality links \textit{as a proportion of external links shown in each session}. Table \ref{tab:low_quality} shows the total counts of low-quality links, the count of all external links, and of all tweets along with low-quality links as a proportion of links vs as a proportion of tweets (averaged across accounts in each group). A mixed effects model for low-quality of links as a proportion of all links indicates significant interaction between the feed and the network-leaning, so we conducted post-hoc comparisons. We did not find evidence for a difference in the proportion of low-quality sources between the algorithmic and chronological feed for right-leaning accounts, and that the algorithmic feed \textit{increased} this proportion for left-leaning ($Z=3.36, p<.01$) and neutral accounts ($Z=6.11, p<.001$).

\begin{table}
\small
\caption{Mean prevalence of links to low-quality information sources, as a percentage of external links vs. as a percentage of all tweets for left-leaning (L), right-leaning (R) and neutral (N) accounts on their algorithmic (Alg) and chronological (Chron) feeds.}
\begin{tabular}{m{.25cm}m{1cm}m{.75cm}m{.75cm}m{.75cm}m{.75cm}m{.75cm}}
 & \textbf{Feed} & \textbf{\% of Links} & \textbf{\% of Tweets} & \textbf{\# of Low-qual} & \textbf{\# of Links} & \textbf{\# of Tweets} \\
\midrule
\multirow{2}{*}{L} & Alg & 9.0\% & 0.1\% & 14 & 157 & 13120 \\
 & Chron & 3.8\% & 0.7\% & 89 & 2412 & 12491 \\
\multirow{2}{*}{N} & Alg & 20.0\% & 1.3\% & 250 & 1237 & 20263 \\
 & Chron & 11.0\% & 2.5\% & 496 & 4708 & 19812 \\
\multirow{2}{*}{R} & Alg & 29.3\% & 0.7\% & 84 & 270 & 12575 \\
 & Chron & 31.5\% & 5.8\% & 708 & 2394 & 12283 \\
\bottomrule
\label{tab:low_quality}
\end{tabular}
\end{table}

\section{Discussion}
 In this study, we investigate the role of Twitter's content recommendation algorithm during the 2022 U.S. midterm elections using automated accounts with varied social networks. 

We find that the recommender has inconsistent effects regarding the promotion of election content. The algorithmic feed decreased the amount of election content shown to left-leaning users, but it did not have a discernible effect for neutral or right-leaning users. In contrast, the recommender increased the visibility of political candidates running for office. This finding extends the findings presented by \citet{Huszar2022-jm} who found that tweets by U.S. elected officials are, on average, amplified by Twitter's algorithm.

Across all network-leaning groups, we find that recommended content came from more right-leaning authors when compared to the chronological feed. For left-leaning accounts, this could be due to the algorithmic feed showing less election content (as found in RQ1), and potentially more apolitical content, which would pull the average partisanship score of post authors toward zero. Additionally, left-leaning accounts had chronological feeds that were significantly more left-leaning than the average of the accounts they follow, so the rightward shift is a correction toward the partisanship of the network (Figure \ref{fig:author_bar}). However, for neutral and right-leaning accounts, the recommended content pushed further away from a politically balanced middle, toward more right-leaning authors. This result complicates recent findings that Twitter's algorithmic feed exposed users to less extreme and less ideologically congruent news sources \cite{Wang2024-zd}, but aligns with a growing number of findings that right-leaning political figures benefit more from algorithmic amplification \cite{Huszar2022-jm, Ye2025-ej}.  

Consistent with previous studies, we find that in general, right-leaning accounts see low-quality information more frequently \cite{Wang2024-zd,Nikolov2021-rh, Gonzalez-Bailon2023-nz}. However, a concerning, and novel finding is the increase in false, misleading, and uncorroborated election rumors on the algorithmic feed for right-leaning accounts and an increase the proportion of links to low-quality domains on the algorithmic feed for left-leaning and neutral accounts. Recent studies on Twitter \cite{Wang2024-zd} and Facebook \cite{Guess2023-ia} found that algorithmic feeds do not affect, or mitigate the prevalence of untrustworthy sources on the respective platforms. Both of these studies use source-level (e.g. domains) labeling, and do not normalize by the frequency of external links. Focusing solely on the volume of low-quality links may overlook other forms of undesirable content (e.g. posts that do not use external links), and mask the mechanisms of how suppressing visibility of \textit{all} external links also suppresses visibility of low-quality links. Furthermore, platforms may be spurred to suppress certain forms of content by research reporting statistics on low-quality links, while neglecting other signals of information quality.   

We find that the chosen social context of the account, network-leaning, is more impactful on exposure to partisan content and election rumors than any detected effect of the algorithm. These results rebut the popular idea of the ``filter bubble", which hypothesizes that algorithmic forces drive users toward increasingly polarized online content \cite{Pariser2011-op}. This finding --- that user preference, rather than algorithmic influence, is the key driver of content exposure --- aligns with a growing body of empirical findings that challenge the filter bubble hypothesis \cite{Ribeiro2023-ti, Hosseinmardi2024-vn, Hosseinmardi2021-nc, Chen2023-tc}. Algorithmic recommendations do not exist in a vacuum, but rather are a function of a user’s specific preferences, the overall content environment, and the mechanism of the algorithm.

\subsection*{Limitations}
This study considers a narrow context --- the 2022 U.S. midterm elections on Twitter --- which may not generalize to other political environments, elections, or platforms. In particular, conceptualizing political preference along a binary axis may not translate well to elections with more than two major parties. Additionally, even within the U.S., the information environment surrounding each election evolves, as do algorithmic designs. As a result, this study can better contribute as a baseline for comparison for evaluating the role of Twitter's content recommender on information visibility during election periods.   

Another limitation is the challenge of deploying and reliably maintaining automated accounts, which can limit the number of audit accounts included in the analysis and which interact with the platform differently than typical users. Yet automated accounts provide the advantage of not only avoiding privacy or ethical implications of a human subjects analysis, but also of an external audit which maintains third party perspective that internally conducted platform studies lack. Further, the audit accounts in this study did not engage with content or other accounts. 
Therefore, outcomes reported in this study isolate the effects of the recommender based on the network of followed accounts.

Finally, the recommendation algorithm studied here is not static and can be changed at any time by the platform without notice to the public \cite{Bagchi2024-hz}. While we do not see evidence of major changes during the course of this study, it is important to keep this in mind when interpreting findings. Additionally, this single platform cannot directly generalize to other platforms, though they may implement similar content recommenders due to similar incentives around user engagement \cite{Narayanan2023-zj}. 

\section*{Conclusion \& Future Work}
In this work we have conducted an algorithmic audit of Twitter's content recommendation system within the context of the 2022 U.S. midterm election. We find that the algorithmic feed influences exposure to election-related content, partisan skew, and the amount of low-quality information, though these effects are small compared to differences across network-leaning groups. Given that platform policies and algorithms are regularly subject to change, the utility of this study lies in the documentation of algorithmic influence at the particular, and meaningful, point in time. Twitter in particular has seen significant changes in content moderation and political discourse following its purchase by Elon Musk in 2022 \cite{Myers2023-al,pew_politics_2024}. Conducted just prior to major change on the platform, this work offers a reference point for future work that quantifies and compares algorithmic influence over time. Continued research into the social implications of algorithmic design are needed, particularly in civic discourse and political engagement.

\bibliography{paperpile, extra}


\onecolumn
\clearpage

\section{Appendix}
\appendix
\label{sec:appendix}

\setcounter{table}{0}
\renewcommand{\thetable}{A\arabic{table}}
\renewcommand{\thesubsection}{\Alph{subsection}}

\subsection{Validation of Election Content}
\label{sec:appendix_election_content}

We conducted a validation of our election-related content labels. To accomplish this, we took a random sample of 200 tweets from our timeline tweets dataset, equally split between those labeled as election-related and those labeled as not related. Two authors manually coded (without access to the dataset labels or one-another's labels) whether each tweet was related to election politics of the 2022 U.S. Midterm election. We then measured our inter-annotator reliability using Cohen's Kappa, both between the human annotators and by comparing each human annotator to the keyword label annotations. At least substantial agreement was reached in all cases, and the level of agreement between each human annotator and the keyword-based annotation was roughly comparable to the agreement between human annotators. The high levels of agreement and comparable performance of human annotators to the keyword method employed validates that our election content annotations are sufficiently accurate for classifying tweets. We report these results below in Tables \ref{tab:appendix_kappas} and \ref{tab:appendix_pct_agreement}. We also provide some examples of tweets from each class in Table \ref{tab:appendix_examples}.

\begin{table}[htp] \centering 
  \caption{Cohen's Kappa scores between human annotators and keyword labels for whether tweets contained election content.} 
  \label{tab:appendix_kappas} 
\begin{tabular}{lll} 
& Labels & Annotator 1 \\ \hline
Annotator 1 & 0.8 & \\
Annotator 2 & 0.68 & 0.79
 
\end{tabular} 
\end{table} 

\begin{table}[htp] \centering 
  \caption{Percent agreement scores between human annotators and keyword labels for whether tweets contained election content.} 
  \label{tab:appendix_pct_agreement} 
\begin{tabular}{lll}
& Labels & Annotator 1 \\ \hline
Annotator 1 & 90\% & \\
Annotator 2 & 84\% & 89\%
 
\end{tabular} 
\end{table} 

\begin{table}[h!] \centering 
  \caption{A sample of five tweets that were classified as election-related and five that were not, to illustrate the kinds of content which were labeled. Tweets were edited to remove identifying tweet links, and unnecessary line breaks.} 
  \label{tab:appendix_examples} 

\renewcommand{\arraystretch}{1.5}
\begin{tabular}{m{.45\linewidth}m{.45\linewidth}}
\textbf{Election-related} & \textbf{Not election-related} \\ \hline
Remember, no matter who wins elections… Jesus will never be removed from office. & Shot my first turkey today!  Scared the crap out of everyone in the frozen food aisle! \\ \hline
Democrats will retain control of Senate as Ron DeSantis’s longtime friend and former roommate loses Nevada Senate race & Not everyone is going to like you. So, strive to become the best version of yourself for internal satisfaction and happiness. Not for some external validation and recognition. \\ \hline
When we show up to vote, we WIN! & Egyptian Museum in Cairo [heart emoji] \\ \hline
Adam Kinzinger just said ``Donald Trump is going to throw Kevin McCarthy under the bus." Let’s the GOP implosion begin! & \#TropicalStormNicole is expected to bring hurricane conditions, including rain and flooding, to parts of Florida’s east coast  beginning late Wednesday. Communities recovering from Hurricane Ian should be cautious. Listen to local officials and follow @FLSERT for updates\\ \hline
 Progressives are stronger than ever and we're ready to fight and deliver for people across our country & Disney just forced out CEO Bob Chapek and is replacing him with Bob Iger, the former CEO, his old boss. Pretty wild.\\ 
\end{tabular} 
\end{table} 

\renewcommand{\thefigure}{A\arabic{figure}}
\setcounter{figure}{0}

\clearpage
\subsection{Media Bias Factcheck}

\begin{figure}[!htb]
     \centering
     \includegraphics[width=.75\linewidth]{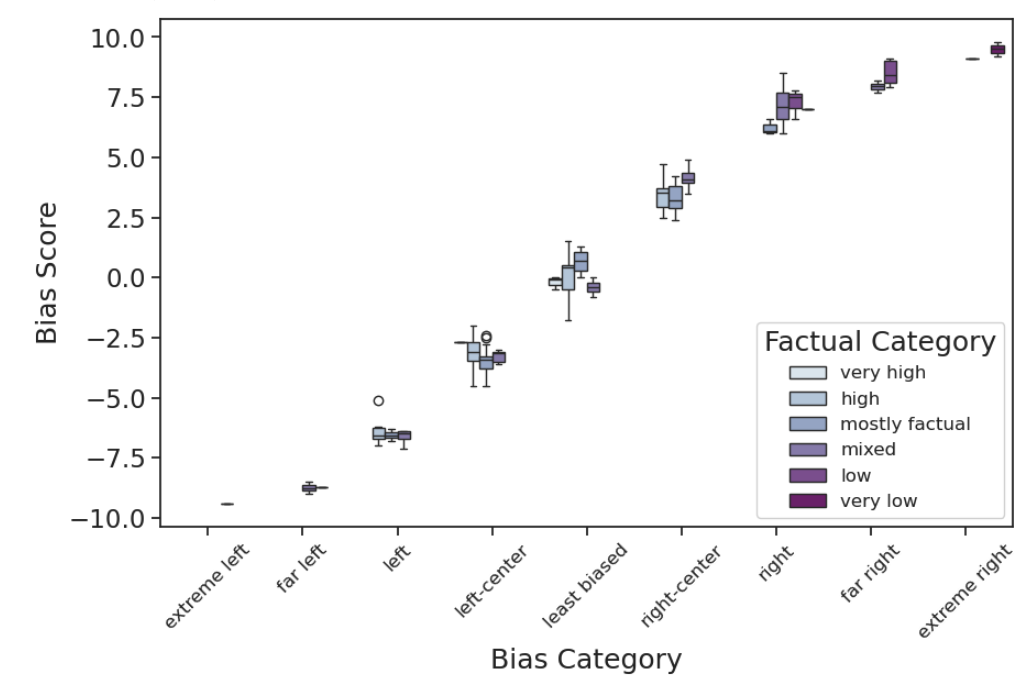}
     \caption{\small Media Bias Factcheck bias scores for the 180 domains that were rated with numerical scores, by bias and factual categories}
\end{figure}

\clearpage
\setcounter{table}{0}

\renewcommand{\thetable}{B\arabic{table}}

\subsection{Regression Model Tables}
\begin{table}[htp] \centering 
  \caption{Election Content Generalized Linear Mixed Model} 
   
\begin{tabular}{@{\extracolsep{5pt}}lD{.}{.}{-3} } 
\\[-1.8ex]\hline 
\hline \\[-1.8ex] 
 & \multicolumn{1}{c}{\textit{Dependent variable:}} \\ 
\cline{2-2} 
\\[-1.8ex] & \multicolumn{1}{c}{keyword\_political} \\ 
\hline \\[-1.8ex] 
 feedalgorithmic & -0.304^{***} \\ 
  & (0.029) \\ 
  network\_leaningN & -1.334^{***} \\ 
  & (0.430) \\ 
  network\_leaningR & 0.022 \\ 
  & (0.477) \\ 
  feedalgorithmic:network\_leaningN & 0.331^{***} \\ 
  & (0.041) \\ 
  feedalgorithmic:network\_leaningR & 0.347^{***} \\ 
  & (0.040) \\ 
  Constant & -0.861^{**} \\ 
  & (0.342) \\ 
 \hline \\[-1.8ex] 
Observations & \multicolumn{1}{c}{1,728} \\ 
\hline 
\hline \\[-1.8ex] 
\textit{Note:}  & \multicolumn{1}{r}{$^{*}$p$<$0.1; $^{**}$p$<$0.05; $^{***}$p$<$0.01} \\ 
\end{tabular} 
\end{table} 

\vspace{2cm}

\begin{table}[!htbp] \centering 
  \caption{Politician Authorship} 
   
\begin{tabular}{@{\extracolsep{5pt}}lD{.}{.}{-3} } 
\\[-1.8ex]\hline 
\hline \\[-1.8ex] 
 & \multicolumn{1}{c}{\textit{Dependent variable:}} \\ 
\cline{2-2} 
\\[-1.8ex] & \multicolumn{1}{c}{politician\_author} \\ 
\hline \\[-1.8ex] 
 feedalgorithmic & 0.288^{***} \\ 
  & (0.069) \\ 
  network\_leaningN & -0.466 \\ 
  & (0.561) \\ 
  network\_leaningR & -0.738 \\ 
  & (0.623) \\ 
  feedalgorithmic:network\_leaningN & 0.356^{***} \\ 
  & (0.084) \\ 
  feedalgorithmic:network\_leaningR & 0.867^{***} \\ 
  & (0.109) \\ 
  Constant & -3.531^{***} \\ 
  & (0.441) \\ 
 \hline \\[-1.8ex] 
Observations & \multicolumn{1}{c}{1,728} \\ 
\hline 
\hline \\[-1.8ex] 
\textit{Note:}  & \multicolumn{1}{r}{$^{*}$p$<$0.1; $^{**}$p$<$0.05; $^{***}$p$<$0.01} \\ 
\end{tabular} 
\end{table}

\begin{table}[htp] \centering 
  \caption{Political Bias Linear Mixed Model} 
   
\begin{tabular}{@{\extracolsep{5pt}}lD{.}{.}{-3} } 
\\[-1.8ex]\hline 
\hline \\[-1.8ex] 
 & \multicolumn{1}{c}{\textit{Dependent variable:}} \\ 
\cline{2-2} 
\\[-1.8ex] & \multicolumn{1}{c}{author\_leaning} \\ 
\hline \\[-1.8ex] 
 feedalgorithmic & 0.120^{***} \\ 
  & (0.014) \\ 
  network\_leaningN & 0.686^{***} \\ 
  & (0.086) \\ 
  network\_leaningR & 1.539^{***} \\ 
  & (0.096) \\ 
  feedalgorithmic:network\_leaningN & -0.011 \\ 
  & (0.017) \\ 
  feedalgorithmic:network\_leaningR & -0.045^{**} \\ 
  & (0.019) \\ 
  Constant & -0.762^{***} \\ 
  & (0.068) \\ 
 \hline \\[-1.8ex] 
Observations & \multicolumn{1}{c}{1,728} \\ 
\hline 
\hline \\[-1.8ex] 
\textit{Note:}  & \multicolumn{1}{r}{$^{*}$p$<$0.1; $^{**}$p$<$0.05; $^{***}$p$<$0.01} \\ 
\end{tabular} 
\end{table}

\begin{table}[!htbp] \centering 
  \caption{MBFC Bias Scores} 
  
\begin{tabular}{@{\extracolsep{5pt}}lD{.}{.}{-3} } 
\\[-1.8ex]\hline 
\hline \\[-1.8ex] 
 & \multicolumn{1}{c}{\textit{Dependent variable:}} \\ 
\cline{2-2} 
\\[-1.8ex] & \multicolumn{1}{c}{mbfc\_bias} \\ 
\hline \\[-1.8ex] 
 feedalgorithmic & 0.543 \\ 
  & (0.402) \\ 
  network\_leaningN & 2.291^{*} \\ 
  & (1.214) \\ 
  network\_leaningR & 7.530^{***} \\ 
  & (1.340) \\ 
  feedalgorithmic:network\_leaningN & 1.745^{***} \\ 
  & (0.485) \\ 
  feedalgorithmic:network\_leaningR & 0.327 \\ 
  & (0.508) \\ 
  Constant & -2.892^{***} \\ 
  & (0.948) \\ 
 \hline \\[-1.8ex] 
Observations & \multicolumn{1}{c}{1,139} \\ 
\hline 
\hline \\[-1.8ex] 
\textit{Note:}  & \multicolumn{1}{r}{$^{*}$p$<$0.1; $^{**}$p$<$0.05; $^{***}$p$<$0.01} \\ 
\end{tabular} 
\end{table} 

\vspace{2cm}

\begin{table}[htbp] \centering 
  \caption{Election Rumors Generalized Linear Mixed Model} 
  
\begin{tabular}{@{\extracolsep{5pt}}lD{.}{.}{-3} } 
\\[-1.8ex]\hline 
\hline \\[-1.8ex] 
 & \multicolumn{1}{c}{\textit{Dependent variable:}} \\ 
\cline{2-2} 
\\[-1.8ex] & \multicolumn{1}{c}{rumor\_tweet} \\ 
\hline \\[-1.8ex] 
 feedalgorithmic & -1.890^{*} \\ 
  & (1.099) \\ 
  network\_leaningN & -0.050 \\ 
  & (0.636) \\ 
  network\_leaningR & 2.207^{***} \\ 
  & (0.595) \\ 
  feedalgorithmic:network\_leaningN & 2.876^{**} \\ 
  & (1.175) \\ 
  feedalgorithmic:network\_leaningR & 3.024^{***} \\ 
  & (1.133) \\ 
  Constant & -4.673^{***} \\ 
  & (0.575) \\ 
 \hline \\[-1.8ex] 
Observations & \multicolumn{1}{c}{1,728} \\ 
\hline 
\hline \\[-1.8ex] 
\textit{Note:}  & \multicolumn{1}{r}{$^{*}$p$<$0.1; $^{**}$p$<$0.05; $^{***}$p$<$0.01} \\ 
\end{tabular} 
\end{table} 

\begin{table}[htbp] \centering 
  \caption{External Links Generalized Linear Mixed Model} 
 
\begin{tabular}{@{\extracolsep{5pt}}lD{.}{.}{-3} } 
\\[-1.8ex]\hline 
\hline \\[-1.8ex] 
 & \multicolumn{1}{c}{\textit{Dependent variable:}} \\ 
\cline{2-2} 
\\[-1.8ex] & \multicolumn{1}{c}{has\_external} \\ 
\hline \\[-1.8ex] 
 network\_leaningN & -0.027 \\ 
  & (0.506) \\ 
  network\_leaningR & 0.051 \\ 
  & (0.561) \\ 
  feedalgorithmic & -3.021^{***} \\ 
  & (0.084) \\ 
  network\_leaningN:feedalgorithmic & 1.153^{***} \\ 
  & (0.092) \\ 
  network\_leaningR:feedalgorithmic & 0.604^{***} \\ 
  & (0.106) \\ 
  Constant & -1.498^{***} \\ 
  & (0.398) \\ 
 \hline \\[-1.8ex] 
Observations & \multicolumn{1}{c}{1,728} \\ 
\hline 
\hline \\[-1.8ex] 
\textit{Note:}  & \multicolumn{1}{r}{$^{*}$p$<$0.1; $^{**}$p$<$0.05; $^{***}$p$<$0.01} \\ 
\end{tabular} 
\end{table} 

\vspace{2cm}

\begin{table}[!htbp] \centering 
  \caption{Low Quality Links Generalized Linear Mixed Model} 
 
\begin{tabular}{@{\extracolsep{5pt}}lD{.}{.}{-3} } 
\\[-1.8ex]\hline 
\hline \\[-1.8ex] 
 & \multicolumn{1}{c}{\textit{Dependent variable:}} \\ 
\cline{2-2} 
\\[-1.8ex] & \multicolumn{1}{c}{low\_quality\_link} \\ 
\hline \\[-1.8ex] 
 feedalgorithmic & 1.048^{***} \\ 
  & (0.312) \\ 
  network\_leaningN & 0.703 \\ 
  & (0.721) \\ 
  network\_leaningR & 2.802^{***} \\ 
  & (0.782) \\ 
  feedalgorithmic:network\_leaningN & -0.494 \\ 
  & (0.325) \\ 
  feedalgorithmic:network\_leaningR & -1.080^{***} \\ 
  & (0.343) \\ 
  Constant & -3.642^{***} \\ 
  & (0.561) \\ 
 \hline \\[-1.8ex] 
Observations & \multicolumn{1}{c}{1,335} \\ 
\hline 
\hline \\[-1.8ex] 
\textit{Note:}  & \multicolumn{1}{r}{$^{*}$p$<$0.1; $^{**}$p$<$0.05; $^{***}$p$<$0.01} \\ 
\end{tabular} 
\end{table}

\end{document}